\documentstyle[12pt]{article}
\def\laq{\raise 0.4 ex \hbox{$<$}\kern -0.8 em\lower 0.62 ex\hbox{$\sim$}}
\def\gaq{\raise 0.4 ex \hbox{$>$}\kern -0.7 em\lower 0.62 ex\hbox{$\sim$}}

\input{epsf}
\def\to{\rightarrow}

\def\AJ{{\it Ap. J.} }
\def\AJL{{\it Ap. J. Lett.} }

\def\APP{{\it Acta Phys. Pol.} }

\def\CQG{{\it Class. Quantum Gravity} }

\def\GRG{{\it Gen. Relativity and Gravitation} }

\def\JHEP{{\it JHEP} }

\def\MPL{{\it Mod. Phys. Lett.} }

\def\NC{{\it Il Nuovo Cimento} }
\def\NP{{\it Nucl. Phys.} }
\def\PL{{\it Phys. Lett.} }
\def\PR{{\it Phys. Rev.} }
\def\PRL{{\it Phys. Rev. Lett.} }

\def\al{\alpha}
\def\be{\beta}
\def\ga{\gamma}
\def\de{\delta}

\def\th{\theta}

\def\ka{\kappa}
\def\la{\lambda}

\def\si{\sigma}

\def\Ga{\Gamma}
\def\De{\Delta}

 \def\frac#1#2{{\textstyle{{#1}\over
{#2}}}} 
\def\lsim{\mathrel{\rlap{\lower4pt\hbox{\hskip1pt$\sim$}}
    \raise1pt\hbox{$<$}}} \def\gsim{\mathrel{\rlap{\lower4pt\hbox{\hskip1pt$\sim$}}
    \raise1pt\hbox{$>$}}}
\def\sqr#1#2{{\vcenter{\vbox{\hrule height.#2pt
         \hbox{\vrule width.#2pt height#1pt \kern#1pt
         \vrule width.#2pt}
         \hrule height.#2pt}}}} \def\square{\mathchoice\sqr66\sqr66\sqr{2.1}3\sqr{1.5}3}

\def\beq{\begin{equation}}
\def\eeq{\end{equation}}
\def\beqa{\begin{eqnarray}}
\def\eeqa{\end{eqnarray}}

\begin{document}

\begin{flushright}
{DF/IST - 1.2003} \\
{May 2004} \\
\end{flushright}
\vglue 1cm

\begin{center}

{{\bf  The Pioneer anomaly in the
context of the braneworld scenario}}

\vglue 0.5cm {O.\ Bertolami and J. P\'aramos}

\vglue 0.2cm

{E-mail addresses: {\tt orfeu@cosmos.ist.utl.pt;
x\_jorge@netcabo.pt}}

\bigskip
{\it Instituto Superior T\'ecnico, Departamento de F\'\i sica,\\}
\smallskip
{\it Av.\ Rovisco Pais, 1049-001 Lisboa, Portugal\\}
\end{center}

\setlength{\baselineskip}{0.7cm}

\vglue 1cm

\date{\today}

\begin{abstract}
We examine the Pioneer anomaly - a reported anomalous acceleration
affecting the Pioneer 10/11, Galileo and Ulysses spacecrafts - in
the context of a braneworld scenario. We show that effects due to
the radion field cannot account for the anomaly, but that a scalar field
with an appropriate potential is able to explain the
phenomena. Implications and features of our solution are analyzed.

\vskip 0.5cm

\end{abstract}


\section{Introduction}
Studies of radiometric data from the Pioneer 10/11, Galileo and
Ulysses have revealed the existence of an anomalous acceleration
on all four spacecrafts, inbound to the Sun and with a (constant)
magnitude of $a_A \simeq (8.5 \pm 1.3) \times 10^{-10}~ m s^{-2}$.
Extensive attempts to explain this phenomena as a result of poor
accounting of thermal and mechanical effects, as well as errors in
the tracking algorithms used, have shown to be unsuccessful
\cite{Nieto}, despite a recent claim otherwise \cite{Scheffer}.

The two Pioneer spacecrafts follow approximate opposite hyperbolic
trajectories away from the Solar System, while Galileo and Ulysses
describe closed orbits. This, together with the fact that the
three designs are geometrically distinct, explains the lack of an
``engineering'' solution for the anomaly. However, it prompts for
a much more intriguing question: what is the fundamental, and
possibly new, physics behind this anomaly?

To answer this, many proposals have been advanced. The range of
ideas is quite diverse and we mention some of them: Yukawa-like or
higher order corrections to the Newtonian potential
\cite{Anderson}; a scalar-tensor extension to the standard
gravitational model \cite{Calchi}; Newtonian gravity as a long
wavelength excitation of a scalar condensate inducing electroweak
symmetry breaking \cite{Consoli}; interaction of the spacecrafts
with a long-range scalar field coupled to gravity \cite{Mbelek,Cadoni}
; an inverse time dependence for the gravitational
constant $G$ \cite{Sidharth}; a length or momentum scale-dependent
cosmological term in the gravitational action functional
\cite{Modanese}; a 5-dimensional cosmological model with a
variable extra dimensional scale-factor and a static external
space \cite{Belayev}; a local Solar System curvature for
light-like geodesics arising from the cosmological expansion
\cite{Rosales}; similarly, a recent work argues that the reported
anomaly is related with the cosmological constant at the scale of
the Solar System \cite{Nottale}, even though this would lead to a
repulsive force; an interaction with ``mirror gas'' or ``mirror
dust'' in the Solar System \cite{Foot}; a superstrong interaction
of photons or massive bodies with the graviton background,
yielding a constant acceleration, proportional to the Hubble
constant \cite{Ivanov}; expansion of solid materials on board deep
space probes and contraction on Earth due to a curved stress field
arising from repetitive tidal action \cite{Guruprasad}; an
expanded PPN-framework so to incorporate a direct effect on local
scales due to the cosmic space-time expansion \cite{Ostvang}; a
result of flavor oscillations of neutrinos in the Brans-Dicke
theory of gravity \cite{Capozziello}; a theory of conformal
gravity with dynamical mass generation, including the Higgs scalar
(capable of reproducing the standard gravitational dynamics and
tests within the Solar System, and yet leaving room for a
Pioneer-like anomaly on small bodies) \cite{Wood}; a gravitational
frequency shift of the radio signals proportional to the distance
to the spacecrafts and the density of dust in the intermediate
medium \cite{Crawford}; resistance of the spacecrafts antennae as
they traverse interplanetary dust \cite{Didon}; a gravitational
acceleration $a \propto r^{-2} $ for a constant $a \gg a_0 =
10^{-10}~ m s^{-2}$ and $a \propto r^{-1}$ for $a \ll a_0 $
\cite{Milgrom}; clustering of dark matter in the form of a
spherical halo of a degenerate gas of heavy neutrinos around the
Sun \cite{Viollier} - amongst a few other suggestions put forward.
It is interesting to mention that in higher-curvature theories of
gravity where the gravitational coupling is asymptotically free
and which have been much discussed in the context of the dark
matter problem \cite{Goldman,Bertolami4,Bertolami5}, a stronger
gravitational coupling is expected on large scales and hence, at
least in principle, to a Pioneer-like anomalous acceleration.

In this work we consider the Pioneer anomaly in the context of the
braneworld scenario. We use the Randall-Sundrum model and
variations to show that gravitational effects such as the one due
to the radion field cannot explain the anomaly. We argue that a
scalar field with a suitable potential implies that geodesics in
this theory exhibit an extra constant attractive acceleration.

\section{Braneworld Theories}

A quite new range of possible scenarios arise in the context of
braneworld theories. In these, one assumes our Universe to be a
3-dimensional world-sheet embedded in a higher dimensional bulk
space. Considerations on the symmetries of the brane and its
topological properties are then taken to constrain the evolution
of matter on the brane and gravity on the brane and in the bulk.

Braneworld theories are a fast developing trend in cosmology
\cite{Randall1, Randall2, Dvali, Gregory2, Kogan}, whose main
feature is to allow for a solution for the hierarchy problem,
whether the typical mass scale of the bulk is comparable with the
electroweak breaking scale, $M_{EW} \sim TeV$.

In this work, we shall consider the Randall-Sundrum braneworld
model and some variations \cite{Randall1, Randall2}. One admits a
scenario with two 3-branes embedded in (4+1)-dimensional space: a
positive tension brane situated at $z=0$ and a negative tension
one at $z=z_c$. The \textit{Ansatz} for the metric takes the form

\beq ds^2 = e^{-2kz} g_{\mu\nu} dx^\mu dx^\nu + dz^2 ~~, \eeq

\noindent which is a solution of the 5-dimensional Einstein's
equations and preserves Poincar\'{e} invariance on each brane. The
constant $k$ is a fundamental quantity of the theory and typically
takes values of order $ k \sim M_{Pl} $, the 4-dimensional Planck
mass, which is dynamically generated from the bulk space Planck
mass $M_5$,

\beq M^2_{Pl} = {M^3_5 \over k} \left(1 - e^{-2kz_c} \right)~~.
\eeq

\noindent This relation is obtained from the derivation of a
4-dimensional effective action by integrating the fifth dimension
away. Notice that $M_{Pl}$ depends weakly on the second membrane
position, in the large $ kz_c $ limit. Physical masses on the
positive brane, however, scale with this distance through the
relationship $ m = e^{-kz_c} m_0 $ (a solution for the hierarchy
problem is obtained for $kz_c \approx 15$). In traditional
compactification schemes, the ``warp'' factor $e^{-2kz}$ in the
metric is absent, and hence integration over the fifth dimension
yields only the volume of the bulk space, $V_n$; for a
$n$-dimensional compact space one obtains $M_{Pl}^2=M^{n+2} V_n$.

\section{Braneworld Scenarios for the Pioneer anomaly}

As a first attempt to explain the Pioneer anomaly within the
context of braneworld theories, one could resort to the appearance
of a tower of Kaluza-Klein massive tensorial perturbations to the
metric. Three problems arise: all gravitons are ordered according
to their mass, so that one cannot freely specify the range of one
of them without affecting the whole tower. Most braneworld models
consider one first light mode with cosmological range, and all
ensuing modes to have sub-millimeter range. It is difficult to
introduce an intermediate scale without abdicating from one of the
two desired extreme cases.

The second problem refers to the fact that any Yukawa
gravitational potential would affect all bodies within range,
independently of their mass (as expected from the Equivalence
Principle). This is in direct contradiction with the lack of an
observed ``anomalous'' acceleration for the planets in the Solar
System.

Thirdly, it has been shown that an Yukawa potential fitted to the
observed effect would reveal the presence of a graviton with range
$L \sim 200~AU$ \textit{and} a negative coupling $\al \sim
-10^{-3}$ \cite{Anderson}. One can adjust the exact values, since
they belong to a solution curve $L = L(\al)$ (the one presented is
a rather ``natural'' choice, since $|\al| \ll 1$ and $L \sim
100~AU$). However, $\al$ is always negative. Since the coupling of
modes is proportional to the normalization factor of their wave
functions in the Kaluza-Klein reduction scheme, this would imply
in a graviton with negative norm - which is unattainable within
current braneworld theories, leading to instabilities at a quantum
level. This latter issue shows that quantized tensor excitations
are not the cause of the Pioneer anomaly.

If one considers instead a single membrane, then the periodic
boundary condition disappears and the modes are no longer
quantized. In the standard Randall-Sundrum scenario
\cite{Randall2}, this amounts to a modification of gravity at
large distances, with the gravitational potential presenting a
$1/r^2$ behaviour, typical of the five-dimensional propagation of
gravity in the bulk. This behaviour also appears in models where
the modes are quantized and dense, and their mass spectrum
approaches a continuous distribution \cite{Kogan}.

A more elaborate model was suggested in Ref. \cite{Gregory2},
which shows a similar behaviour of the gravitational potential. At
large distances it also goes as $1/r^2$, while at intermediate
distances it contains a repulsive logarithmic term. In both cases
(and more exotic ones, see Refs. \cite{Kogan, Rubakov} and
references therein), a continuous spectrum of Kaluza-Klein
excitations can be shown to be unable of explaining the Pioneer
effect.

An alternative solution could be that the Pioneer anomaly reflects
the influence of the radion field $f(x)$, a scalar perturbation of
the metric corresponding to translational zero modes, related to
relative motion of the two branes. Since its moduli is usually
stabilized due to some \textit{ad hoc} potential whose effect is
superimposed on the usual warped metric (see Ref.
\cite{Goldberger} and references therein), one could view the
radion as an additional field on each brane. We discuss its effect
on a test particle and verify to which extent it could provide an
explanation for the Pioneer anomaly.

Following Ref. \cite{Gregory}, the induced metric on the positive
($z=0$) brane which includes this perturbation is, in Gaussian
coordinates ($g_{zz}=-1$ and $g_{z_\mu}=0$),

\beq h_{\mu\nu}^+= \eta_{\mu\nu}[1-2kf(x)] + {1 \over
2k}f_{,\mu\nu}~~~, \label{metric} \eeq

\noindent where $f(x)$ is constrained by the requirement that in
the vacuum

\beq \square^2 f = 0~~~, \label{square}\eeq

\noindent and

\beq \square^2 \equiv \left( \eta^{\la \si} f_{;\la}
\right)_{;\si} = \eta^{\la \si} f_{,\la \si} - \eta^{\la \si}
\Ga_{\la \si}^\al f_{,\al} \eeq

\noindent is the 4-dimensional d'Alembertian. This arises from the
Israel junction conditions imposed on each brane so to ensure the
$\mathcal{Z}_2 $ symmetry. The metric $\eta^{\la \si}$ is the
unperturbed one on the positive brane. In the presence of matter,
Eq. (\ref{square}) is not homogeneous and is related to the trace
of the energy-momentum tensor of the matter distribution.

We now search for a solution $f(r)$ that is both static and
spherically symmetric. We work in spherical coordinates where $f'$
denotes $f_{,r}$, and $f_{,\th} = f_{,\phi} = f_{,t} = 0 $. In the
weak field limit, one has $\eta_{00}= 1 + 2U$ and $ \eta_{rr} =
-(1-2U) $, where $U= - GM_\odot/r \equiv -C / r $ is the Newtonian
gravitational potential; in units where $ c = \hbar = 1 $, $ C =
1.5 ~km $. Thus, one obtains

\beq \square^2 f = - f'' - {2 \over r} f' = 0 ~~, \eeq

\noindent neglecting the term due to the curvature, which is
proportional to $ U'(r) f'(r) $.

A simple solution is given by $ f(r) =k^{-1} (A / r + B) $, with
$B \ll 1$ a dimensionless constant, $[A]=L$ and $A/r \ll 1$ within
the desired range. Hence, the induced metric becomes

\beqa h_{00}^+ = [1+2U(r)][1-2kf(r)] \nonumber \\
h_{rr}^+ = -[1 - 2U ][1-2 k f(r)] + k^{-1} f''(r)~~. \eeqa

We next proceed by considering the spacecrafts as point particles.
Thus, they follow time-like geodesics of the obtained vacuum
metric. The acceleration is then

\beqa {a^r} & = & -\Ga_{00}^r = {1 \over 2} h^{rr}\partial_r
h_{00} =  -{1 \over 2} [1 + 2U] \left[{1 \over 1-2kf - k^{-1}
f''(r)}
\right] \partial_r\left[(1 + 2U) ( 1 - 2 kf ) \right] \nonumber \\
& \simeq & - [1 + 2U] [1 + 2kf + k^{-1} f''] \partial_r \left[U -
kf -2kUf) \right] \nonumber \\ & = & -{C + A \over r^2} + {-2A^2 +
4AC +2C^2 \over r^3} + {4A(3A-C)C\over r^4} \\ \nonumber & & -
{2A(A+C+8AC^2k^2) \over k^2 r^5} + {4AC(3A+C) \over k^2 r^6} -
{16A^2C^2 \over k^2 r^7} ~~, \label{accel} \eeqa

\noindent and one can see that no constant term arises, rendering
this approach unsuitable to account for the discussed anomalous
acceleration.

Another possible explanation for the Pioneer anomaly could be
related with a bimetric theory exhibiting Lorentz symmetry
breaking. A proposal along these lines was discussed in Ref.
\cite{BertolamiM} and is briefly presented in the Appendix A.

In principle, a bimetric theory could arise from induced effects
on the 3-brane of gravity in the bulk. However, as shown in Ref.
\cite{Bandos}, Goldstone modes resulting from the spontaneous
breaking of coordinate diffeomorphisms carry no extra degrees of
freedom due to the positions of the branes in curved spacetime,
but rather manifest themselves in the form of an extra field in
the energy-momentum tensor, involving the induced metric $
\ga_{mn} = \partial_m x^\mu \partial_n x^\nu g_{\mu\nu}(x) $.
Thus, although invariance under general coordinate transformations
does not hold along the fifth dimension, it is a symmetry on each
brane. Since the resulting singular energy-momentum tensor cannot
be treated as a ``mass'' term for the graviton (at least at the
linear level), massive gravitons do not arise due to the
interaction of the brane with gravity in the bulk. It is argued
that this is crucial to preserve the $ r^{-2} $ long-range
behaviour of gravity.

Nevertheless, it should be pointed out that long-range
modifications to Newton's law have been the focus of many
braneworld related proposals aiming to explain the accelerated
expansion of the Universe (see e.g. Ref. \cite{Dvali} and
references therein). It is important to realize that the above
arguments are valid in the absence of topological obstructions on
the brane. Hence, no extra degrees of freedom arise
when the metric is taken as a dynamic variable under local,
``small'', diffeomorphisms. However, under large gauge
transformations \cite{Adawi}, where the gauge parameters assume
different values in non-connected asymptotic regions and have
different global behaviour, solitonic-type solutions are admitted.

These solutions are deformations, $h_{\mu\nu} $, of a particular
metric solution $ g^{(1)}_{\mu\nu}$ of the Einstein's equations
with a source term given by the extra field arising from the
breaking of diffeomorphism invariance along the fifth direction,
such that the deformed metric $g^{(1)}_{\mu\nu} + h_{\mu\nu} $ is
a solution of the same equation. We have verified that known
solitonic-type solutions do not give rise to a constant
acceleration term \cite{Stojkovic, Cembranos} that could explain
the Pioneer anomaly.

In what follows, we shall derive the general behaviour of
a scalar field with a potential and calculate its effect on the
motion of a test particle.

\section{Scalar Field Coupled to Gravity}

As a possible explanation for the Pioneer anomaly, we consider the
effect induced by the presence of a scalar field $ \phi$  with
dynamics driven by a potential $ V(\phi) \propto - \phi^{-\al}(r)
$, with $\al > 0$. The form of this potential closely resembles
that of some supergravity inspired quintessence models
\cite{Linde, Binetruy, Brax}. In the context of the braneworld
scenario, the quintessence potential has the form $V \propto
\phi^{-\al}(t) $, with $2 < \al < 6$ \cite{Maeda}. Notice that, in
this proposal, there is a spatial and not a time dependence. Also,
the sign of our potential is reversed, so to yield a static
\textit{attractive} acceleration.

As usual, we assume a small perturbation to the Minkowsky metric
and solve it in terms of the energy-momentum tensor of the field.

The metric can be written as

\beq
g_{\mu\nu} = \eta_{\mu\nu} + h_{\mu\nu}
\label{pertmetric}
\eeq

\noindent where in the weak field limit (in spherical coordinates)

\beq (\eta)_{\mu\nu} = diag(1 + 2 U(r), -1+2U(r), -r^2, -r^2 sin^2
\th)~~, \eeq

\noindent and

\beq
h_{\mu\nu} = diag(f(r),-h(r),-h(r) r^2,-h(r) r^2sin^2 \th)~~.
\eeq

\noindent Notice that the bimetric character arises from the
assumption that the field $\phi$ expresses the effect of the
induced metric arising from the spontaneously broken
diffeomorphisms in the curved spacetime.

The Lagrangian density of the static scalar field takes the form

\beq
{\cal L}_\phi = {1 \over 2} \eta_{\mu\nu} \partial_\mu \phi
\partial^\nu \phi - V(\phi) = {1 \over 2} \eta_{rr}\left(\phi
'\right)^2 + A^2 \phi^{-\al} ~~,
\label{lagrangian}
\eeq

\noindent where $A$ is a constant. The scalar field obeys the
equation of motion

\beq
\square^2 \phi+ {d V(\phi) \over d \phi} = 0 ~~,
\label{motion}
\eeq

\noindent which yields, neglecting a term proportional to $U'(r)
\phi'(r)$

\beq
\phi''(r) + {2 \over r} \phi'(r) = \al A^2 \phi^{-\al -1} ~~,
\label{eqmotion}
\eeq

\noindent and admits the solution

\beq
\phi(r) = \left( (2 + \al) \sqrt{\al \over 8 + 2 \al} A r
\right)^{2 \over 2 + \al} \equiv \be^{-1} r^{2 \over 2 + \al}~~.
\label{solution}
\eeq

\noindent
Notice that this solution is singular at $r=0$. Its regularization is discussed in Appendix B.
Thus, in terms of $r$, the potential and gradient terms
are given by

\beq
V(\phi(r)) = -A^2  \be^\al r^{-{2 \al \over 2 + \al}}~~,
\eeq

\noindent and

\beq
{1 \over 2} \left(\phi ' (r) \right)^2 = A^2 \left({\al \over
4 + \al}\right)  \be^\al r^{-{2 \al \over 2 + \al}} = -\left({\al
\over 4 + \al}\right) V(\phi(r)) ~~.
\eeq

The Lagrangian density is given in the Newtonian limit by

\beq
{\cal L}_\phi = -{4 \over 4 + \al}V(\phi(r)) = {4 A^2 \over 4 +
\al} \be^\al r^{-{2 \al \over 2 + \al}}~~.
\label{lagrangian1}
\eeq

The energy-momentum tensor of the scalar field is obtained by the
expression

\beq
T_{\mu\nu} = \partial_\mu \phi \partial_\nu \phi -
\eta_{\mu\nu} {\cal L}_\phi ~~,
\label{stress}
\eeq

\noindent so that its components are given by

\beqa
T_{00} & = & -\eta_{00} {\cal L}_\phi \nonumber =
-[1 + 2 U(r)] {4 A^2 \over 4 + \al} \be^\al r^{-{2 \al \over 2 + \al}} \nonumber \\
T_{rr} & = & \phi'(r)^2 - \eta_{rr} {\cal L}_\phi = (2 + \al) {2
A^2 \over 4 + \al}\be^\al r^{-{2 \al \over 2 + \al}} \nonumber \\
T_{\th\th} & = & - \eta_{\th \th} {\cal L}_\phi = r^2 {4 A^2
\over 4 + \al} \be^\al r^{-{2 \al \over 2 + \al}}\nonumber \\
T_{\varphi\varphi} & = & - \eta_{\varphi\varphi} {\cal L}_\phi = r^2
sin^2 \th {4 A^2 \over 4 + \al} \be^\al r^{-{2 \al \over 2 + \al}}
~~, \eeqa

\noindent where we have assumed that the spatial perturbation to
the metric is very small, $h(r) \ll 1$. The trace is given by

\beq T \equiv T_\al^\al = \eta^{\mu\nu} T_{\mu\nu} = - (8 + \al){2
A^2 \over 4 + \al} \be^\al r^{-{2 \al \over 2 + \al}} ~~.
\label{trace} \eeq

We now turn to the linearized Einstein's equations:

\beq {1 \over 2} \nabla^2 h_{\mu\nu} = \ka\left( T_{\mu\nu} - {1
\over 2} \eta_{\mu\nu} T \right) ~~, \label{fieldeqs} \eeq

\noindent where $\ka = 8 \pi G $. The $0-0$ component is

\beq f''(r) + {2 \over r} f'(r) =  2\ka \left( 1 - {2 C \over r}
\right) A^2  \be^\al r^{-{2 \al \over 2 + \al}} \label{00} \eeq

\noindent whose solution is, for $\al \neq 2 $,

\beq f(r) = (2 + \al)^2 A^2 \ka \be^\al r^{-{2 \al \over 2 + \al}}
\left({C r\over \al -2} + {r^2 \over 12 + 2\al}\right)~~, \eeq

\noindent and, for $\al = 2$,

\beq f(r) = \sqrt{3 \over 2} {A \ka r \over 2} - \sqrt{6} AC \ka
~log \left({r \over 1~m} \right) ~~.\eeq

\noindent
Notice that we have dropped the homogeneous solution of Eq. (\ref{00}) of
the form $1/r$ since it can be absorbed by the Newtonian term.
Therefore, we obtain for the acceleration

\beq a_r = -{C \over r^2} + (2 +\al)A^2 \ka \be^a r^{-{2 \al \over
2 + \al}} \left( {C \over 2} - {r \over 6 + \al} \right) ~~.
\label{alfa} \eeq

\noindent For $\al = 2$, it reads

\beq a_r = -{C \over r^2} - \sqrt{3 \over 2} {A \ka \over 4} +
\sqrt{3 \over 2} {A C \ka \over r} ~~. \label{alfa2} \eeq

Eqs. (\ref{alfa}) and (\ref{alfa2}) indicate that a constant
anomalous acceleration exists only for $\al = 2$. The first term
is the Newtonian contribution, and one can identify the constant
term with the anomalous acceleration $a_A = 8.5 \times 10^{-10}~ m
s^{-2} $, setting

\beq A =  4 \sqrt{2 \over 3} {a_A \over \ka} = 4.7 \times
10^{42}~m^{-3}~~. \eeq

The remaining term, proportional to $r^{-1}$, is much smaller than
the anomalous acceleration for $ 4C/r \ll 1$, that is, for $r \gg
6~km$; it is also much smaller than the Newtonian acceleration for
$r \ll ~ 2.9 \times 10^{22}~km \approx 100~Mpc$, clearly covering
the desired range. It can be shown that higher order terms affect
the third term in Eq. (\ref{alfa2}) and yield a negligible
correction to the first term.

For consistency, we now show that $h(r) \ll 1$. For that, we use
the $r-r$ component of Eq. (\ref{fieldeqs}) with $\al = 2$:

\beq h''(r) + {2 \over r} h'(r) = -{A \ka\over \sqrt{6} r}~~,
\label{rr} \eeq

\noindent whose solution is $ h(r) = -A \ka r / 2 \sqrt{6}$. This
term is negligible for $r \ll 5000~ Mpc$, confirming the validity
of the approximation. One can improve the previous result of Eq.
(\ref{alfa2}) to first order in $h(r)$:

\beq a^{(1)} =  {a_A \over 1 + h(r)} \simeq a_A (1 - h(r)) = - {C
\over r^2} + {5 \over 6} \sqrt{3 \over 2} {A C \ka \over r} -
\sqrt{3 \over 2} {A \ka \over 4} \left(1 - \sqrt{2 \over 3} A C
\ka \right) - \left({A \ka \over 4}\right)^2 r ~, \eeq

\noindent which, asides from the small correction to $a_A$ and a
$5/6$ factor in the $r^{-1}$ term, introduces a linear term which
is negligible for $r \ll 5 \times 10^{29}~Mpc $.

Thus, we see that an anomalous acceleration is a clear prediction
of our model within the Solar System. Hence, an hypothetical
dedicated probe to confirm the Pioneer anomaly \cite{Bertolami,
Anderson2} does not need to venture into too deep space to detect
such an anomalous acceleration, but just to a distance where it is
measurable against the regular acceleration and the solar
radiative pressure, actually approximately from Saturn onwards.

It can be shown that the potential $V(\phi) \propto -\phi^{-2}$ is
the only one that can account for the anomalous acceleration.
Indeed, from Eq. (\ref{00}) we can write in the same approximation

\beq {1 \over r^2} [r^2 f'(r)]' = \chi \ka V(r) \eta_{00}(r)~~,
\eeq

\noindent where $\chi$ is a dimensionless factor depending on the
potential ($\chi = - 2$ for $\al = 2$). Hence,

\beq f'(r) = {\chi \ka \over r^2} \int r^2V(r) \eta_{00}(r)~~.
\eeq

Since, from Eq. (\ref{alfa2}), the anomalous acceleration is given
by

\beq a_r = -{1 \over 2}f'(r) = -{\chi \ka \over 2 r^2} \int V(r)
r^2 \eta_{00}(r)~~, \eeq

\noindent and $ \eta_{00} = 1- 2 C / r $, one concludes that the
potential must have the form $V(r) = V(\phi(r)) = -A / r $ or
$V(r) = B $, where $B$ is a constant. The constant potential
solution also provides an anomalous acceleration, but yields a
conflicting linear term:

\beq f_{(const)} = -2 B C \ka r + B \ka r^2 /3 \eeq

\noindent and hence

\beq a_{r~(const)} = -{C \over r^2} - BC \ka + {B \ka r \over
3}~~. \eeq

By identifying the constant term with the anomalous acceleration,
one immediately finds $ B = 0.03 ~kg~m^{-3}$; as a result, the
linear term dominates the constant acceleration for $ r \gg
1.5~km$. Hence, we consider the case $\chi = -2$ or $\al = 2$.

Indeed, the equation of motion (\ref{motion}) can be written as

\beq \nabla^2 \phi = {1 \over r^2} [r^2 \phi'(r)]' = {V'(r) \over
\phi'(r)} ~~, \eeq

\noindent and therefore

\beq \phi'(r) [r^2 \phi'(r)]' = A~~. \eeq

\noindent Thus, the only real solution of this differential
equation is given, up to a constant factor, by $ \sqrt{r} $.
Hence, $ r \propto \phi^2(r) $ and $V(\phi) \propto
-\phi^{-2}(r)$, as argued.

With $\al=2$, the energy of this scalar field grows
logarithmically. This can be seen as a manifestation of a global
symmetry breaking, as in the case of global cosmic strings, where
the same behaviour is found \cite{Nielsen}.

Notice that these results apply only to the case of bodies with
negligible mass, such as the considered spacecrafts. In the case
of planets, endowed in general with rotation and spin, their
dynamics are described by the solution of Eq. (\ref{fieldeqs})
with the appropriate energy-momentum tensor $T_{\mu\nu} =
T^\phi_{\mu\nu} + T^{planet}_{\mu\nu}$, with $T^\phi_{\mu\nu} \ll
T^{planet}_{\mu\nu}$. Hence, no anomalous acceleration is expected
for celestial objects.

Furthermore, in what concerns the nature of the source of the scalar field, we assume
it is the Sun and that matter couples with the scalar field in the following way:

\beq
{\cal L}_{int} = \phi \sum_i f_i  \bar{\psi}_i \psi_i~~,
\label{int_l}
\eeq
where, $\psi_i$ stands for different fermion species and
$f_i$ are couplings constants.

We now look at a possible breaking of the Equivalence Principle.
It can be shown (see \cite{Carrol, Prep} and references therein) that changes to
geodesic motion arise (to first order) from spatial variations of
mass, proportionally to $m'(r)/m(r)$. Hence, one must look
at changes induced on the mass of a test particle. These will
occur if the scalar field acquires a vacuum expectation value,
such as in the standard case of the Higgs mechanism, or in
the cosmologically relevant cases (\cite{Carrol}).

In our proposal, the potential for the scalar field is
negative and monotonically increasing. Hence, the
effective potential $V_{eff}(\phi) =
V(\phi) + \phi \sum_i f_i n_i $, where $n_i$ is the density of
different matter species coupled to the scalar field,
does not develop a minimum. Therefore, no mass changes as well as no
violations of the Equivalence Principle are expected.

Finally, we look at the propagation time delay induced by this
bimetric theory. If it is not negligible, then the value of the
anomalous acceleration should be corrected, since the radiometric
data which supports it is based on the Doppler effect
\cite{Nieto}.

For simplicity, we consider only the worst-case scenario: a test
particle travelling at a constant velocity of $v = 10^{-5}c $
(approximately the current velocity of Pioneer 10/11), following a
linear path away from the Sun and at a distance $r_0$ from it. If
the light signal is emitted with a proper period $T_0$, its
relation with the period recorded on Earth, $T$, is

\beqa T & = & \int_{T_0} \sqrt{[1+ f(r(t))] - v^2}~ dt \simeq
\int_{T_0} \left[ \sqrt{1-v^2} + {f(r) \over 2\sqrt{1-v^2}}
\right] ~{dt \over v} \nonumber \\ & = & \int_{\De r} \left[
\sqrt{1-v^2} + \sqrt{3 \over 2} A \ka~ {\left[ r/4 - C ~ log
\left({ r \over 1~m } \right) \right] \over \sqrt{1-v^2}} \right]
~{dr \over v} ~~,\eeqa

\noindent where $\De r = v T_0$. This yields

\beq T = \left( \sqrt{1-v^2} + \sqrt{3 \over2} {A \ka r \over 4
\sqrt{1-v^2}} \right) T_0 ~~, \eeq

\noindent where we have used that $r \gg 1.5~km$ and $r \gg \De
r$. Hence, the bimetric effects on the time delay can be safely
disregarded for $r \ll 3000~Mpc$.

\section{Conclusions and Outlook}

In this paper we have investigated the possibility of explaining
the Pioneer anomaly within the framework of braneworld scenarios.
We have eliminated both ``new'' tensorial (massive gravitons) as
well as scalar (radion) degrees of freedom as candidates for a
solution. We found that the anomalous acceleration could be due to
the presence of a negative potential scalar field, with a
potential $V \propto -\phi^{-2}(r) $ similar to some supergravity
inspired quintessence models.

Notice that the approach considered here, contrary to naive
thinking, has no implications for the puzzle of the rotation curve
of the galaxy. Indeed, assuming a galaxy to be virialized, one can
describe the rotation of a layer at a distance $r$ of the galactic
core as $ v^2(r) = G M(r) /r $, where $M(r)$ is the total mass
inside the layer. Observation shows that $v^2(r)$ displays a
steady rise until a threshold of about 10 kpc ($\sim 2 \times 10^9
AU$), and a constant plateau from there (see eg. Ref.
\cite{Trimble}). This leads one to model $M_{Gal}(r) \sim r$ and
hence to postulate the presence of dark matter.

Starting from Eq. (\ref{alfa2}), we can derive a different
expression for $v^2(r)$:

\beq v^2(r) = {C^* \over r} - \sqrt{3 \over 2} C^* \ka + \sqrt{3
\over 2} {A^* \ka r\over 4}~~, \label{alt_rotational} \eeq

\noindent where the superscript $^*$ refers to galactic values.
This curve does not describe the observed data and would lead one
to model $ M(r) $ as

\beq M(r) = M_{Gal}(r) \left[ 1 + \sqrt{3 \over 2} A^* \left(\ka r
+ {\ka r^2 \over 4 C^*} \right) \right] \label{Mr}~~. \eeq

This clearly differs from usual dark matter models due to the
higher order terms in Eq. (\ref{Mr}). However, likewise the case
of planets, the test masses here cannot be viewed as point-like
objects, and hence have to be treated with its energy-momentum
tensor. Therefore, we are lead to conclude that the origin of the
flattening of the rotation curves of galaxies does not have its
roots in the induced bimetric theory.

\section{Appendices}

\subsection{Appendix A.}

A bimetric theory of gravity was first proposed by Rosen
\cite{Rosen}. Its action is given by

\beq
S = {1 \over 64 \pi G} \int d^4x \sqrt{-\eta}
[~\eta^{\mu\nu}g^{\al\be}g^{\ga\de} (g_{\al\ga |\mu}g_{\be\de
|\nu} - {1 \over 2} g_{\al\be |\mu}g_{\ga\de|\nu}) +
{\cal L}_{M}(g_{\mu\nu})]~~,
\label{action}
\eeq

\noindent where the vertical line $|$ denotes covariant derivation
with respect to the background metric $\eta_{\mu\nu}$ only, and
${\cal L}_M$ is the matter Lagrangian density. The resulting
equation for the dynamical gravitational field is given by:

\beq
\square_\eta g_{\mu\nu} - g^{\al\be}\eta^{\ga\de}g_{\mu\al |
\ga}g_{\nu\be |\ga} = -2 \ka (g/\eta)^{1/2}(T_{\mu\nu}-{1 \over 2}
g_{\mu\nu} T)~~,
\label{fieldeqsbi}
\eeq

\noindent
where $T \equiv T_{\mu\nu} g^{\mu\nu}$ and
$\square_\eta$ is the d'Alembertian operator with respect to
$\eta_{\mu\nu}$. Notice that the momentum-energy tensor couples
only to the dynamical metric $g_{\mu\nu}$. We can always choose
coordinates in which $\left( \eta_{\mu\nu} \right) =
diag(-1,1,1,1)$, the Minkowsky metric, and $ \left( g_{\mu\nu}
\right) = diag(-c_0,c_1,c_1,c_1)$, where $c_0$ and $c_1$ are
parameters that may vary on a Hubble $H^{-1}$ timescale
\cite{Rosen}.

This theory explicitly breaks Lorentz invariance. This is better
understood by resorting to the Parametrized Post-Newtonian (PPN)
formalism: a systematic expansion of first-order $1/c^2$ terms in
the Newtonian gravitational potential and related quantities
\cite{Will1}. It turns out that any metric theory of gravitation
can be classified according to ten PPN parameters:
$\ga,\be,\xi,\al_1,\al_2,\al_3,\zeta_1,\zeta_2,\zeta_3,\zeta_4$.
These are the linear coefficients of each possible first-order
term (generated from rest mass, energy, pressure and velocity),
and relate a particular theory with fundamental aspects of
physics: conservation of linear and angular momentum,
preferred-frame and preferred-location effects, nonlinearity and
space-curvature per unit mass, etc.

Einstein's General Relativity, the most successful theory up to
date, exhibits a set of PPN parameters with $ \be = \ga = 1$, the
remaining being equal to zero. Rosen's bimetric theory has $ \be =
\ga = 1 $, but a non-vanishing $\al_2 = c_0/c_1 - 1$ coefficient.
This indicates that the theory is semi-conservative (it lacks
angular momentum conservation) and exhibits preferred-frame
effects: Lorentz invariance is broken and the Strong Equivalence
Principle does not hold. It is worth pointing out that the
breaking of Lorentz invariance has been recently very much
discussed. Indeed, possible signatures of the breaking of this
symmetry arise from ultra-high energy cosmic rays with energies
beyond the Greisen-Zatsepin-Kuzmin cut-off, $E_{GZK} \simeq 4
\times 10^{19}~eV$, (see Ref. \cite{Bertolami1} for a discussion
on the astrophysical aspects of the problem), from the observation
of gamma radiation from faraway sources with energies beyond
$20~TeV$, and from the longitudinal evolution of air showers
created by ultra-high energy hadronic particles and which seem to
imply that pions are more stable than expected (for an update see
\cite{BertolamiM, Bertolami2} and references therein). Of course,
Lorentz invariance holds with great accuracy as observed
deviations are quite small, $\delta < 3 \times 10^{-22}$
\cite{Lamoreaux} from direct measurements, and even smaller from
the study of ultra-high energy cosmic rays $\delta \simeq 1.7
\times 10^{-25}$ \cite{Coleman,Bertolami3}.

By linearizing Eq. (\ref{fieldeqsbi}) in the vacuum, one obtains
the wave equations for weak gravitational waves, whose solution is
a wave propagating with speed $c_g = \sqrt{c_1/c_0}$. Thus,
$\al_2$ measures the relative difference in speed (measured by an
observer at rest in the Universe rest frame) between
electromagnetic and gravitational waves, inducing a time delay in
the propagation of light signals \cite{Will1}. This effect has
been claimed to be put under test in a recent experiment using the
close celestial alignment of Jupiter and the quasar $J0842+1835$.
This analysis yield $ c_g / c = 1.06 \pm 0.21$ \cite{Kopeikin},
corresponding to $ \al_2 = - 0.11 \pm 0.35 $. Note, however, that
this result is still controversial, and an alternative
interpretation suggests that it sets instead a limit on $\ga$ and
$\al_1$, actually $\ga=1$ and $\al_1=0$, being unrelated to the
velocity of propagation of gravity \cite{Will2}.

A rigorous study of deviations between the Sun's spin axis and the
ecliptic has led to the experimental constraint $|\al_2| < 1.2
\times 10^{-7} $ \cite{Nordtvedt}. Improving this bound as well as
finding new means of verifying its implications are clearly of key
importance. Interestingly improvements on the measurement of Sun's
oblateness (and ensuing spin) as well as on the PPN parameters
$\beta$, $\gamma$ and the combination $\eta \equiv 2 -\beta + 2
\gamma$ are on the list of objectives of the ambitious BepiColombo
mission to Mercury \cite{Balogh}.

Due to their small mass, self-gravitation is also absent for the
considered spacecrafts. Thus, these can be regarded as particles,
which enables us to calculate their acceleration by simply
computing the time-like geodesics of the full metric, $h_{\mu\nu} =
\eta_{\mu\nu} + g_{\mu\nu} $.

In the weak field limit, $v \ll c$, one has

\beq
{a^i} \simeq -\Ga_{00}^i = {1 \over 2} h^{i \la}
\partial_\la h_{00} ~~,
\label{accel_bi}
\eeq

\noindent from which one can identify the radial anomalous
component of the acceleration:

\beq
\vec{a}_A = c_1 \vec{\nabla}_r U - {(1 - c_1) \over 2}~
\vec{\nabla}_r c_0 ~~.
\label{anomalous}
\eeq

It can be immediately seen that, if $c_0$ and $c_1$ are
homogeneous in space, the derived anomalous acceleration is not
constant, which contradicts the observation. Therefore, we assume
these parameters depend on the distance to the Sun, that is,
$c_0=c_0(r), ~ c_1=c_1(r)$. Given the constraint

\beq
\al_2 = \left|{c_0(r) \over c_1(r)} -1 \right|  < 4 \times
10^{-7} ~~,
\eeq

\noindent we assume that both coefficients have the same
$r$-dependence, so that $\al_2$ is homogeneous. We consider the
choice

\beq
c_0=D~r~~,~~c_1=F~r  ~~,
\label{ansatz}
\eeq

\noindent with $ D, F > 0 $ so that the resulting anomalous
acceleration is inbound. According to Ref. \cite{Nordtvedt},
$|D/F-1| < 4 \times 10^{-7} $ and hence $ D \simeq F $. Note that
$D$ cannot be exactly equal to $F$, since this implies that
$\al_2=0$.

Substituting {\it Ansatz} Eq. (\ref{ansatz}) into Eq.
(\ref{anomalous}), we find

\beq
a_A = -{D \over 2} \left[ 1 - {F \over D}{2 C \over r} - F r
\right] \simeq -{D \over 2} \left[1 - {2 C \over r} - D r \right]~~.
\eeq

\vspace{0.3cm}

\noindent
Hence, $ D = 2 a_A = 1.9 \times 10^{-26}~ m^{-1}$ and we
see that the distance-dependent contributions to $a_A$ are
negligible for $ r $ lying in the interval

\beq
[2 C, D^{-1}] = [3~km, 3.5 \times 10^{14}~ AU]~~,
\label{interval}
\eeq

\noindent
which is consistent with the fundamental assumption that
$c_0, c_1 \ll 1 $.

Unfortunately, this elegant solution for the Pioneer anomaly
cannot be taken seriously, given the behaviour of Rosen's theory
in what concerns gravitational waves. Indeed, in Ref.
\cite{Will3}, it is argued that Rosen's bimetric theory is
fundamentally flawed, since it predicts dipole gravitational
radiation beyond the limits measured from the pulsar $ PSR ~ 1913
+ 16 $ \cite{Taylor}. Moreover, the solution proposed by Rosen of
considering a combination of retarded and advanced gravitational
waves \cite{Rosen2} implies in contradiction with the observed
quadrupole gravitational radiation from binary pulsars
\cite{Taylor}. Thus, Rosen's bimetric theory cannot be considered
a viable theory of gravity.

\subsection{Appendix B.}

Equation (\ref{eqmotion}) is not
satisfied at the origin, the center of the Sun,
as solution of Eq.(\ref{solution}) is singular at $r=0$.
However, this solution can be regularized in the standard way
by introducing a source term in the Lagrangian
density (\ref{lagrangian}). This can be performed by considering a thin shell
model, dividing the space into two regions, $r \ge r_0$
and $r<r_0$, with $r_0$ a positive constant. Assuming that the Pioneer anomaly
is verifiable everywhere in the Solar System, it follows
that $r_0$ is smaller than $R_\odot$, the Sun's radius - this
assumption can be relaxed, since it is currently not accessible to
experiment, due to the suppression of the effect by the larger
solar wind and Newtonian terms.

Naturally, one assumes that the solution of Eq.(\ref{solution}),
$\phi(r) \equiv \phi(r)_+$, is valid for $r \ge r_0$. For $r <
r_0$, we consider the solution

\beq
\phi(r) = \phi_-(r) \equiv C + {\alpha \over 6} A^2
C^{-\alpha - 1}r^2 + {\cal O}\left(r^4\right)~~,
\eeq

\noindent
where $C$ is a regularization constant.
Continuity of the scalar field at $r=r_0$ implies that C is a
solution of $\phi_-(r_0) = \phi_+(r_0)$. Even though the full solution
is now regular at $r=0$, the derivative
changes value at $r=r_0$ as
$\phi_+'\left(r_0\right) \neq \phi_-'\left(r_0\right)$
and, as before, Eq.(\ref{eqmotion}) is not satisfied at $r=r_0$. A suitable
solution requires adding to Eq. (\ref{eqmotion}) the
term
$\left(\phi_+'\left(r_0\right)-\phi_-'\left(r_0\right)\right)\delta\left(r-r_0\right)$,
which demands an additional term in the Lagrangian density Eq.
(\ref{lagrangian}):

\beq
{\cal L}_\phi\to {\cal L}_\phi + \phi
\left(\phi_+'\left(r_0\right) -
\phi_-'\left(t_0\right)\right)\delta\left(r-r_0\right)~~.
\label{source_l}
\eeq

\noindent
Hence, one can remove the ``inner'' solution $\phi_-(r)$ by taking
the $r_0 \rightarrow 0$ limit and keeping the source term in
Eq.(\ref{source_l}).

\vspace{0.3cm}

\noindent
{\bf Acknowledgments}

\vspace{0.2cm}

\noindent The authors wish to thank Cl\'ovis de Matos, Michael
Martin Nieto, Andreas Rathke and Martin Tajmar for useful
discussions on the Pioneer anomaly. This research was partially
developed while the authors were attending the Third COSLAB
Workshop, in Bilbao; we are grateful to the staff of the
University of the Basque Country for their hospitality. JP is
sponsored by the Funda\c{c}\~{a}o para a Ci\^{e}ncia e Tecnologia
(Portuguese Agency) under the grant BD~6207/2001.


\end{document}